\documentclass{revtex4}
\usepackage{graphicx}
\usepackage{fancyhdr}
\usepackage{amsmath}
\usepackage{color}
\usepackage{subfigure}
\usepackage{epstopdf}
\usepackage{epsfig}
\usepackage{amssymb}
\usepackage{bm}
\usepackage{bbm}
\newcommand{\beq}{\begin{equation}}
\newcommand{\eeq}{\end{equation}}
\newcommand{\ov}{\overline}

\begin{document}

\title{Top FB asymmetry vs. (semi)leptonic B decays in multi-Higgs-doublet models}

%

\author{P. Ko}
\affiliation{School of Physics, KIAS, Seoul 130-722, Korea}

\author{Yuji Omura}
\affiliation{Physik Department T30, Technische Universit\"{a}t M\"{u}nchen,
James-Franck-Stra$\beta$e, 85748 Garching, Germany}

\author{Chaehyun Yu}
\affiliation{School of Physics, KIAS, Seoul 130-722, Korea}

\begin{abstract}

We investigate
the semileptonic and leptonic $B$ decays, $B \to D^{(*)} \tau \nu$ and 
$B \to \tau \nu$ in our multi-Higgs-doublet models, 
while keeping in mind the enhancement of the top forward-backward (FB) 
asymmetry ($A^t_{\rm FB}$) at the Tevatron.
In these models, the extra (pseudo)scalar bosons contribute 
to $A^t_{\rm FB}$ with large flavor-changing Yukawa couplings 
involving the top quark, but  
the contribution of the charged Higgs boson to $A^t_{\rm FB}$ is negligible. 
However, it may significantly affect $B$ decays: especially, 
$B\to D^{(*)}\tau \nu$ and $B \to \tau \nu$.   
We investigate the constraints on the $B$ decays, 
based on the recent results in BaBar and Belle experiments, 
and discuss the possibility that the allowed parameter 
region in the $B$ decays can achieve large $A^t_{\rm FB}$.

\end{abstract}

\maketitle

\thispagestyle{fancy}


\section{Introduction}

In Ref.~\cite{Ko-Top}, the present authors proposed flavor models 
with extra Higgs fields, which are gauged by U(1)' symmetry. 
The models slightly break the criteria of Ref.~\cite{Ko-2HDM} 
by assigning flavor-dependent U(1)' charges
to the right-handed (RH) up-type quarks.
In these models, the potentially problematic flavor-changing neutral-currents
(FCNCs) for $B$-$\bar{B}$ and $K$-$\bar{K}$ mixings are forbidden, but
certain amounts of FCNCs could still be allowed while keeping the consistency
with experimental data.
For example, neutral CP-even and CP-odd scalars can have large 
$(t,q)_{q=u,c,t}$ elements
of Yukawa couplings, which may enhance the top forward-backward (FB)
asymmetry ($A^t_{\rm FB}$) at the Tevatron. 
The strong constraints from LHC can be accommodated thanks to 
the destructive interference among the scalars~\cite{Ko-Top,Ko-LHC}.
The charged Higgs boson, which is necessarily present 
in the multi-Higgs-doublet models, does not have sizable contribution
to the top quark production at hadron colliders, but may have large couplings
to the bottom quark because of theoretical consequences.
This indicates that the models could strongly be constrained by $B$ physics.
For example, the $(b,u)$ element of the charged Higgs, which 
has strong correlation with the $(t,u)$ coupling of the pseudoscalar boson,
is constrained by $B\to \tau \nu$. 
Similarly, the large $(b,c)$ element of the charged Higgs corresponding
to the large $(t,c)$ coupling of the pseudoscalar boson could significantly
modify the branching ratios for the semileptonic $B$ decays, 
$B\to D^{(*)} \tau \nu$.

Recently, the BaBar collaboration announced interesting experimental results for
the semileptonic $B$ decays, $B\to D^{(*)} \tau \nu$~\cite{Lees:2012xj}.
The ratios $R(D^{(*)})$ defined by the branching ratios for 
$B\to D^{(\ast)}\tau \nu$ to those for $B\to D^{(\ast)}l \nu$ ($l=e,\mu$),
\begin{equation}
R(D^{(*)}) = \textrm{BR}(B \to D^{(*)} \tau \nu)/
\textrm{BR}(B \to D^{(*)} l \nu),
\end{equation}
are $R(D)= 0.440 \pm 0.072$ and $R(D^*)=0.332 \pm 0.030$ 
at BaBar~\cite{Lees:2012xj}.
These values deviate from the Standard Model (SM) predictions, 
$R(D)_\textrm{SM}=0.297 \pm 0.017$ \cite{Kamenik:2008tj} and 
$R(D^*)_\textrm{SM}=0.252 \pm 0.003$ \cite{Fajfer:2012vx},
by $2.2\sigma$ and $2.7\sigma$, respectively~\cite{Crivellin:2012ye}.
Combining $R(D)$ and $R(D^*)$, one finds that the discrepancy becomes
about $3.4 \sigma$~\cite{Lees:2012xj}, which   
might be an evidence of new physics. 
(The recent Lattice calculation relaxes the deviation
in $R(D)$ \cite{Bailey:2012jg}.) Motivated by this BaBar discrepancy,
new physics scenarios are widely discussed and
one good candidate for new physics is a charged Higgs boson in the extended 
SM with extra Higgs doublets~\cite{Fajfer:2012jt, Crivellin:2012ye}.

The average value of the branching ratio for the leptonic $B$ decay, 
$B\to \tau \nu$, which was measured at BaBar~\cite{Aubert:2008zzb} 
and Bell~\cite{Hara:2010dk}, is 
$\textrm{BR}(B \to \tau \nu)=(1.67 \pm 0.3) \times 10^{-4}$~\cite{Asner:2010qj}.
Although there are some uncertainties from $|V_{ub}|$ and the $B$ meson 
decay constant $f_B$ in the SM calculation,
this measurement is slightly inconsistent with the SM prediction,
for example, $\textrm{BR}(B \to \tau \nu)_\textrm{SM}= 
( 0.84 \pm 0.11 ) \times 10^{-4}$ 
given by the UTfit Collaboration~\cite{Bona:2009cj}.
However, the recent Belle measurement 
$\textrm{BR}(B\to \tau\nu)= 0.72^{+0.27}_{-0.25}\pm 0.11$ by making use of
a hadronic tagging method for tau decays 
with the full data sample~\cite{Adachi:2012mm} is consistent
with the SM prediction and the combination of the two is also consistent
within uncertainties.
Since the same new physics (e.g., a charged Higgs boson) may affect
both the semileptonic and leptonic $B$ decays,
combined analysis of the $B$ decays will strongly constrain
such a new physics scenario.

In Ref.~\cite{Ko:2012sv}, the authors investigated 
if the multi-Higgs-doublet models can explain 
the discrepancies in the $B$ semileptonic decays while keeping consistency
with $\textrm{BR}(B\to \tau \nu)$. Then they discussed the possibility that
the allowed parameter region can achieve the large enough $A^t_\textrm{FB}$ 
which was observed at the Tevatron. In this paper, we summarize the results 
and more explicitly estimate the upper bound on
the enhancement of $A^t_\textrm{FB}$ in the scalar-exchanging scenario.

The remainder of this paper is organized as follows.
In Sec.~\ref{sec:Setup}, we give a brief review 
of our models proposed in Ref.~\cite{Ko-Top}. We assign flavor-dependent
$U(1)'$ charges to the RH up-type quarks and introduce extra $U(1)'$-charged 
Higgs doublets in order to write Yukawa couplings for the up-type quarks including 
top quark. In Sec. \ref{sec:top FB}, we discuss the top FB asymmetry in our models, and
show the favored parameters to enhance $A^t_\textrm{FB}$.
Sec. \ref{sec:B} is devoted to the constraints on especially our two-Higgs-doublet model (2HDM) from 
the $B$ decays. Then, we summarize our results in Sec.~\ref{sec:summary}.

\section{Models with extra Higgs and gauged flavor $\bm{U(1)'}$ }
\label{sec:Setup}
One of the interesting extensions of the SM is to add extra Higgs doublets
to the SM, for examples, two-Higgs-doublet model (2HDM).
In general, such an extension generates the FCNC problem at tree level
through neutral Higgs mediation unless each up- and down-type quark
couples only with one Higgs doublet.
This flavor problem can be avoided by controlling the flavor
structures of Yukawa couplings. A simple way is to assign
a new symmetry to extra Higgs and matter fields.
The most popular symmetry is to introduce
an extra $Z_2$ symmetry with soft breaking terms~\cite{Glashow}.    
In Refs.~\cite{Ko-Top} and~\cite{Ko-2HDM}, 
the authors proposed gauged $U(1)'$ symmetry  
to the SM fermions instead of $Z_2$ symmetry.
Then extra Higgs doublets charged under $U(1)'$ should be introduced
to generate proper mass terms of SM fermions charged under $U(1)'$.
Especially, in Ref.~\cite{Ko-Top}, models where only RH up-type quarks are 
charged flavor-dependently under $U(1)'$ was proposed. 
Then, only FCNCs involving the top quark can be enhanced.  
In these models, interactions between the Higgs doublets charged under $U(1)'$ 
and the RH up-type quarks are given in the form
\begin{equation}
\label{eq:yukawa}
V_y= y^u_{ij} \overline{Q_i} \widetilde{H_j} U_{Rj} 
+y^d_{ij} \overline{Q_i} H_2 D_{Rj} + y^l_{ij} 
\overline{L_i}  H_2 E_{Rj}  + h.c. .
\end{equation}
Here $H_j (j=u,c,t)$ are charged under $U(1)'$. 
$H_2$ has the same quantum numbers as the SM Higgs doublet.
The other fermions in the SM interact with $H_2$. 

According to Eq.~(\ref{eq:yukawa}), there may, in general, be up to four Higgs 
doublets: $H_2$ and $H_{u,c,t}$, but the number of Higgs doublets 
depend on the $U(1)'$ charge assignment. In Ref.~\cite{Ko-Top},
the authors constructed both 2HDM and three-Higgs-doublet model (3HDM)  
by assuming the $U(1)'$ charge assignments
$(u_j)=(0,0,1)$ and $(u_j)=(-1,0,1)$, respectively.
In the 2HDM we identify $H_u$ and $H_c$ as $H_2$ and $H_t$ as $H_1$,
while in the 3HDM, $H_u$ as $H_1$, $H_c$ as $H_2$, and $H_t$ as $H_3$, 
respectively.

In Ref.~\cite{Ko-Top}, the Yukawa couplings of (pseudo)scalars and charged Higgs are calculated explicitly, and
we notice that the form of Yukawa couplings in Eq. (\ref{eq:yukawa}) realizes large $(t,q)_{q=u,c,t}$ Yukawa couplings because of the top mass. The other elements are suppressed by the light quark masses.
We can expect that the large $(t,u)$ and $(t,c)$ couplings of (pseudo)scalars enhance $A^t_{\rm FB}$ and achieve 
the BaBar discrepancy.

\section{Top FB asymmetry in multi-Higgs-doublet models}
\label{sec:top FB}
The top FB asymmetry at the Tevatron is one of the interesting signals 
in the top-quark sector because it might be an evidence of new physics.
The recent updated result of $A^t_{\rm FB}$ at CDF 
in the lepton+jets channel with data of 
a luminosity of $8.7$ fb$^{-1}$  is
$A_\textrm{FB}^t=0.162\pm0.047$~\cite{cdfnew},
which is consistent with the previous measurements at CDF and D0.
While the SM predictions are $0.072^{+0.011}_{-0.007}$ 
at the next-to-leading order (NLO) + next-to-next-to-leading logarithm
accuracies~\cite{smnlo,smnlo2} and $0.087\pm 0.010$ with NLO corrections
for the electroweak interactions~\cite{smewnlo,smewnlo2}, respectively.  
There is about 2$\sigma$ deviation between the SM predictions
and the experimental results in the integrated $A_{\textrm{FB}}^t$ 
at the Tevatron.

\begin{figure}[h]
\centering
\includegraphics[width=80mm]{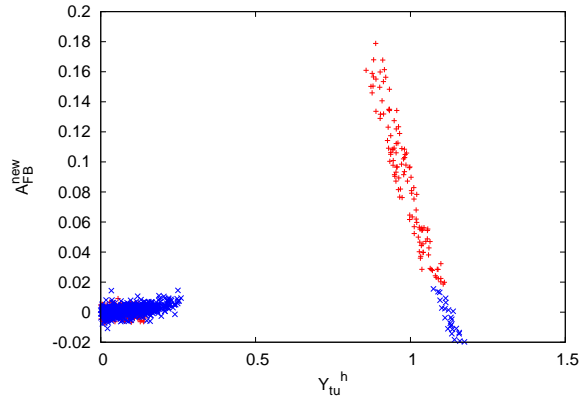}
\caption{The enhancement of $A_\textrm{FB}^t$ from new physics 
$((A_\textrm{FB}^t)^{\rm new})$  vs. $(t,u)$-element of Yukawa coupling. 
The red points are for $m_a < 200$ GeV and the blue points are for 
$m_a \geq 200$ GeV. 
All points are consistent with the $t\overline{t}$ total cross section 
at the Tevatron.} \label{figure-FB}
\end{figure}

In our models, (pseudo)scalars and $Z'$ have large FCNCs between $u$ and $t$,
and their $t$-channel exchanges in the $u \overline{u} \to t \overline{t}$
process can enhance $A_\textrm{FB}^t$ at the Tevatron.
The mediators are neutral particles so that they also allow the same-sign
top-quark pair production process, 
$u u \to t t$, and the scenarios have been tested by the process 
at the LHC~\cite{same-sign, same-sign2}.
As pointed out in Ref. \cite{same-sign}, only $Z'$ scenario, which corresponds to 
the case that the scalar exchanging contributions are negligible, has been excluded. 
Recently, the CMS collaboration updated the upper bound on 
the cross section for the same-sign top-quark pair production:
$\sigma_{tt} <0.3$pb \cite{same-sign3}. 
\begin{figure}[h]
\centering
\includegraphics[width=80mm]{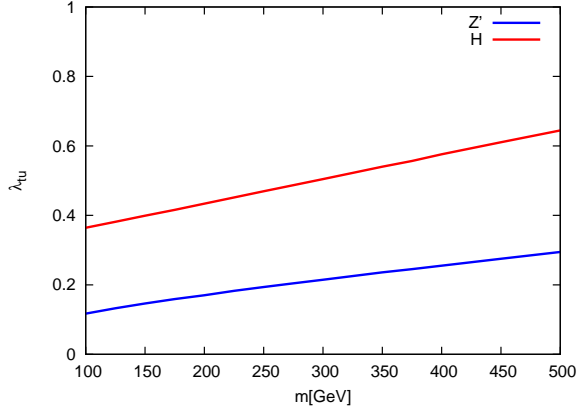}
\caption{Bound on $(t,u)$ couplings of (pseudo)scalars and $Z'$ 
from the same-sign top-quark pair production~\cite{same-sign3}. 
 } \label{figure-samesign}
\end{figure}
In Fig.~\ref{figure-samesign}, the bounds on $(t,u)$ couplings of 
(pseudo)scalars and $Z'$ are described. 
The blue line is only $Z'$ case with the interaction lagrangian 
$\lambda_{tu} Z'^{\mu} \overline{t_R} \gamma_{\mu} u_R$ and the $Z'$ mass $m$
while the red line is only one (pseudo)scalar $h(a)$ case with
the interaction lagrangian $\lambda_{tu} h(a) \overline{t_L} u_R$ and 
the (pseudo)scalar mass $m$.
In the one scalar case, $\lambda_{tu}$ should be so small that we cannot 
achieve the large enhancement of $A_\textrm{FB}^t$ \cite{Ko-Top}. 
As suggested in Ref. \cite{Ko-Top, Ko-LHC}, the bound from the same-sign top-quark pair production
can be relaxed
if destructive interferences work among the contributions from each mediator. Especially, in the case that one CP-even scalar and 
one pseudoscalar have the same $(t,u)$ couplings and the same masses, 
the contributions of the scalars totally cancel each other, 
and $\sigma^{tt}$ becomes zero, if the contributions of the other mediators 
to $\sigma^{tt}$ are negligible. 
The enhancement of $A_\textrm{FB}^t$ in the degenerate case is explicitly 
estimated in Fig.~\ref{figure-FB}. 
$Y^h_{tu}$ is the $(t,u)$ coupling of the CP-even scalar and the pseudoscalar, 
and the (pseudo)scalar mass $m_a$ is heavier (lighter) than $200$ GeV 
for the blue (red) points. 
We see that $Y^h_{tu}$ should be around one 
and $m_a$ should be as small as possible.
If the mass is too light compared with top quark mass, the branching ratio 
for the top quark decaying to $Wb$ will be modified.
This means that the additional enhancement from the scalar exchanging is at most $0.03$, 
when the mass is around $200$ GeV and the coupling is around $1$.
\section{(semi)leptonic B decays in multi-Higgs-doublet models}
\label{sec:B}
Based on the result in the previous section, 
we discuss the possibility that the enhancement of $A_\textrm{FB}^t$ 
is compatible with the (semi)leptonic $B$ decays
in BaBar and Belle experiments.

\begin{figure}[!t]
\begin{center}
{\epsfig{figure=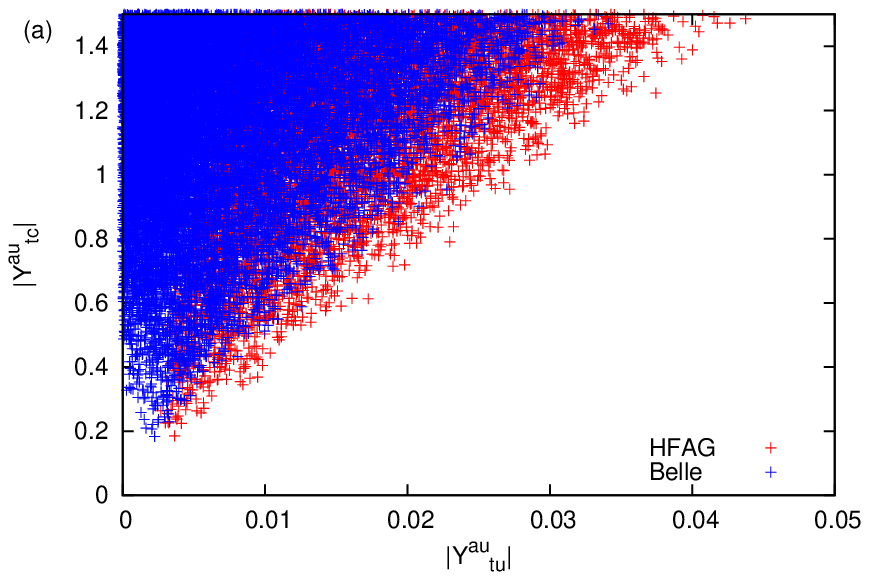,width=0.5\textwidth}}{\epsfig{figure=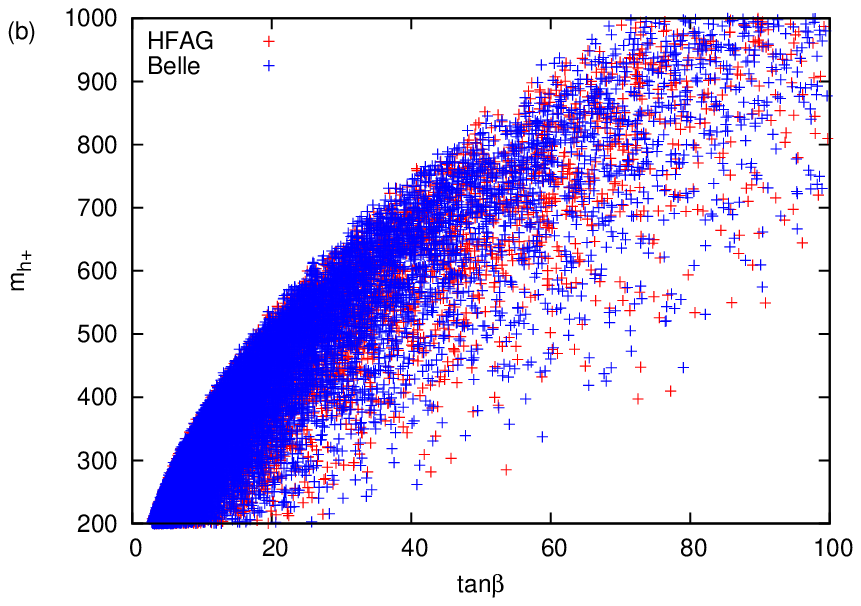,width=0.5\textwidth}}
\end{center}
\vspace{-0.5cm}
\caption{
Bounds on (a) $|Y^{au}_{tu}|$ and $|Y^{au}_{tc}|$ and
(b) $\tan \beta$ and $m_{h^+}$ in 2HDM. We used the relation $Y^{-u}_{bq}=\sqrt{2} V^{*}_{tb} Y^{au}_{tq}(q=u,c)$, ignoring the other elements of the Yukawa coupling.
The points are consistent with $R(D^{(*)})$ within $1 \sigma$. 
The red points are consistent with the combined data
of $\textrm{BR}(B\to \tau \nu)$~\cite{Asner:2010qj}
while the blue points are in agreement with the new Belle data
of $\textrm{BR}(B\to \tau \nu)$~\cite{Adachi:2012mm}.
}
\label{fig:2HDM-Y}
\end{figure}

Only charged Higgs boson contributes 
to the $B$ decays at the tree level, but there are theoretical relations
of Yukawa couplings and masses between the pseudoscalar and charged Higgs bosons:
$Y_{ij}^{-u} = \sum_l V^*_{li} Y^{au}_{lj} \sqrt{2}$ and 
$m^2_{h^{+}}=m^2_a -  \widetilde{\lambda}_{12} \frac{ v^2}{2}$ in our 2HDM,
where $V_{ij}$ is the CKM matrix and $ \widetilde{\lambda}_{12}$ 
is a dimensionless coupling in the Higgs potential.
$Y^{au}_{ij}$ and $Y_{ij}^{-u}$ are the Yukawa couplings of 
the pseudoscalar and charged Higgs $(h^+)$ in the mass bases:
$ Y^{au}_{ij} a \ov{u_{Li}} u_{Rj}$ and $Y^{-u}_{ij} h^- \ov{d_{Li}} u_{Rj}$.
That is, the mass and coupling of the pseudoscalar can 
indirectly be constrained by the B decays.
In Fig.~\ref{fig:2HDM-Y} (a) and (b), we show favored regions in our 2HDM
(a) for Yukawa couplings $|Y^{au}_{tu}|$ and $|Y^{au}_{tc}|$, 
and (b) $\tan\beta$ and $m_{h^+}$, which are consistent with $R(D)$ 
and $R(D^\ast)$ at BaBar within $1\sigma$, respectively. 
The constraint from $D^0$-$\ov{D^0}$ mixing at the one-loop level is also imposed.
The red points are consistent with the combined data for
$\textrm{BR}(B\to \tau\nu)$ while the blue points agree with
the recent Belle data for $\textrm{BR}(B\to \tau\nu)$.
$|Y^{au}_{tu}|$ is restricted to be less than $0.05$
while $|Y^{au}_{tc}|$ is allowed to be $O(1)$. 
In order to account for the discrepancies in $R(D^{(*)})$,
the Yukawa coupling $|Y^{au}_{tc}|$ has to be sizable 
and its lower bound is about $0.2$. 
We can conclude that it is impossible that we can achieve large $A^t_{\rm FB}$ in 2HDM
to be consistent with the $B$ decays.
If the $R(D^{(*)})$ converge to the SM prediction in the future, we would not need consider such  a large new physics effect and a large $Y^{au}_{tc}$, so that we 
would be able to choose the points with a large $Y^{au}_{tu}$. 

In Ref. \cite{Ko:2012sv}, the authors also discussed the three-Higgs-doublet model (3HDM) as the solution
to achieve the enhancement $A^t_{\rm FB}$
and the BaBar discrepancy. 
In fact, both deviations requires large $(t,u)$ and $(t,c)$ couplings, but 
the constraint from $D^0$-$\ov{D^0}$ mixing at the one-loop level also limits the parameter space strongly.
However, we could find the allowed region in Ref.~\cite{Ko:2012sv}.

\section{summary}
\label{sec:summary}
In this paper, we investigated the semileptonic and leptonic 
$B$ decays in our 2HDM, which were proposed 
in Refs.~\cite{Ko-Top,Ko-LHC} in order to account for
the $A^t_{\rm FB}$ measured at the Tevatron.  
Then we discussed the possibility that the enhancement 
of the $A^t_{\rm FB}$ is compatible with the constraints from the $B$ decays. 

Our model predicts large $(t,q)_{q=u,c,t}$ Yukawa couplings of (pseudo)scalars 
and naturally realizes large $A^t_{\rm FB}$ 
through the $t$-channel (pseudo)scalar exchanges.
In our multi-Higgs-doublet models, not only CP-even scalar boson but also 
pseudoscalar boson are necessary to have $O(1)$ $(t,u)$-element Yukawa 
couplings in order to achieve the large $A^t_{\rm FB}$  
while keeping the consistency with the strong bound 
from the same-sign top-quark pair production 
signal at the LHC~\cite{Ko-Top,Ko-LHC}. 
If the pseudoscalar and the CP-even scalar masses are degenerate and the couplings are also same,
the additional enhancement from the scalar exchange 
is at most $(A^t_{\rm FB})^{\rm new} \sim 0.03$ 
when the mass is around $200$GeV and the coupling is around $1$.
 
On the other hand, such a large  $(t,u)$ element of the pseudoscalar Yukawa 
coupling allows a large $(b,u)$ element to appear  in the charged 
Higgs Yukawa couplings. This implies that our models may 
predict large deviations from the SM predictions in $B$ physics.

The recent BaBar results of $R(D^{(*)})$ seem to suggest large flavor changing couplings between $b$ and $c$,
and they may be evidences of new physics beyond the SM. 
One of the good candidates is a charged Higgs boson
and the $(b,c)$ element of the charged Higgs Yukawa coupling, 
which corresponds to large $Y^{au}_{tc}$ in the 2HDM,
should be $O(1)$ to explain the discrepancies.
On the other hand, the experimental results of $B \to \tau \nu$, 
which constrain the $(b,u)$ element of the Yukawa couplings, 
are becoming consistent
with the SM. This implies that the charged Higgs contribution to
this decay should be small.
That is, $Y^{au}_{tu}$ should be small but 
$Y^{au}_{tc}$ should be large.
In fact, we could not find the points  in Fig.~\ref{fig:2HDM-Y} (a) with large 
$Y^{au}_{tu}$ which is needed for large $A^t_{\rm FB}$.
If the $R(D^{(*)})$ converge to the SM prediction in the future, 
we would not need consider such  a large new physics effect 
and a large $Y^{au}_{tc}$, so that we would be able to choose the points 
with a large $Y^{au}_{tu}$ as discussed in Ref.~\cite{Ko:2012sv}. 
However, it may be difficult to consider a light pseudoscalar mass 
$(\sim 200)$ GeV in the 2HDM
because of the bound from $B \to \tau \nu$.
As a solution to realize such large $Y^{au}_{tu}$, the 3HDM is proposed 
in Ref.~\cite{Ko:2012sv}, and 
we could find the favored parameter points which achieve large 
$A^t_{\rm FB}$ and the BaBar discrepancy
without the conflict with the leptonic $B$ decay.




\begin{acknowledgments}
We thank Korea Institute for Advanced Study for providing computing resources (KIAS
Center for Advanced Computation Abacus System) for this work.
This work is supported in part by Basic Science Research Program through the
National Research Foundation of Korea (NRF) funded by the Ministry of Education Science
and Technology 2011-0022996 (CY), by NRF Research Grant 2012R1A2A1A01006053 (PK
and CY), and by SRC program of NRF funded by MEST (20120001176)
through Korea Neutrino Research Center at Seoul National University (PK).
The work of YO is financially supported by the ERC Advanced Grant project “FLAVOUR” (267104).
\end{acknowledgments}

\bigskip 

\end{document}